\documentstyle[12pt]{article}

\begin{document}
\baselineskip=15pt
\title{ Deformation of Nuclei Close to the Two-Neutron Drip Line 
 in Mg Region }
\author{ 
 J.~Terasaki, H.~Flocard \\
 Division de Physique Th\'eorique\thanks{Unit\'e de recherches des
 Universit\'es Paris XI et Paris VI associ\'ee au CNRS.},
 Institut de Physique Nucl\'eaire, \\
  F-91406 Orsay Cedex, France
 \and
         P.-H.~Heenen\thanks{Directeur de Recherches FNRS.} \\
 Service de Physique Nucl\'eaire Th\'eorique,\\
 U.L.B.-C.P.229, B-1050 Brussels, Belgium
 \and
  P.~Bonche \\
SPhT
\thanks{Laboratoire de la DSM} -CEA Saclay,F- 91191
 Gif sur Yvette Cedex, France }
\maketitle

\begin{abstract}
We present Hartree-Fock-Bogoliubov (HFB) calculations 
of the ground states of even Mg isotopes. A Skyrme force 
is used in the mean field channel
and a density-dependent zero-range  force in the pairing channel. 
 $^{40}$Mg and  $^{20}$Mg are predicted to be at the two-neutron 
and two proton drip-lines respectively.
A detailed study of the quadrupole deformation properties
of  all the  isotopes shows
that the ground states of $^{36,38,40}$Mg are strongly deformed
with significantly different deformations for the neutrons and protons.
Our study supports the disappearance of the $N$ = 28 
shell gap in the Mg and Si isotopes.
\end{abstract}



\section{Introduction}
 The existence of deformed drip-line nuclei has been
recently questionned \cite{Ta95} 
on the basis of a detailed analysis 
of the available data.
Tanihata et al. have argued that with few exceptions,
the last bound isotopes of a chain could be related to
a sub-shell closure in a spherical shell model.
The number of cases supporting their analysis
is however limited
since the neutron drip line is known only
up to Z=9. 
A justification relying on results of
fully microscopic
calculations is not easy to obtain. For evident reasons of simplicity,
most studies of drip-line 
nuclei  have been restricted to
spherical shapes \cite{Do94,Do96,Do96b,Hi91,Sm93}. 
>From their outcomes, it has been suggested that the single particle gaps
associated with magic numbers  significantly decrease 
near the drip lines \cite{Do94}. 
Recently, two recent studies of deformation in the Mg\cite{Re96}
and Si-Ca\cite{We96} regions in which
pairing correlations were described within the
constant gap BCS approximation have found sizable deformations 
for some drip line nuclei.
However, no final conclusions can be drawn 
from these studies since the BCS approximation is known to break down when
bound and unbound single-particle states interact 
via  pairing correlations\cite{Do84}. 

In a recent publication\cite{Te96},
we have presented a method based on the solution 
of the Hartree-Fock-Bogoliubov (HFB) 
equations on a 3-dimensional cartesian mesh. It 
describes consistently both the
pairing correlations and the asymptotic behaviour of
the wave functions of weakly bound systems.
In this calculation, the effective interaction is 
considered separately in two different channels:
a Skyrme force to describe the mean-field (particle-hole channel) and   
a density-dependent zero-range force for the pairing correlations
(particle-particle channel).
Tests of the method on the even Ni isotopes 
have shown that it accounts correctly
for the interactions between bound and continuum
single-particle states.
Since all the Ni isotopes are spherical,
the analysis of the results were easier because of
the several degeneracies of 
single particle and quasi-particle (qp)
energies. On the other hand,
the potential of investigation
of the method is broader than shown by the
application presented in ref.\cite{Te96}.
Indeed, it can as well handle deformed nuclei
since all single particle states are calculated within a 
three-dimensional geometry
and allows the description of triaxilly deformed nuclei.

This work is mostly devoted to a study of Mg isotopes.
This isotopic chain has been selected because
large deformations have been predicted for 
the heaviest isotopes
on the basis of the BCS approximation.
Also, together with neighboring isotopic chains, 
these nuclei have been extensively studied 
experimentally.  
Many experimental works have looked for
deformation effects 
at $N = 20$  in the $A = 30$ mass region. 
Results on isotope shifts of Na isotopes\cite{Hu78,To82}
are consistent with an appearance of deformations at $N = 20$ 
although not yet fully conclusive 
because the measure of the quadrupole moment 
of $^{31}$Na is still out
of reach experimentally. 
More convincing evidence comes from measurements of 
masses in $^{31,32}$Mg \cite{De83}, 
$^{30-32}$Na and $^{31-33}$Mg\cite{Gi87}, although
is not exempt of ambiguities\cite{Vi86}. 
However, the flattening of the $S_{\rm 2n}$ curve
in Mg isotopes around $N$ = 20 appears now to be
well established \cite{Wa93}.  
Systematics  of the energies 
of the first 2$^+$ states \cite{De79}
may also be interpreted as a sign of
deformation in $^{32}$Mg.
{}Finally, a large $B(E2)$ value\cite{Mo95}
has recently been measured for the 
transition between the first 2$^+$ state and 
the ground state in $^{32}$Mg. 
Most experimental data point towards the existence of a spherical
shell closure at $N$ = 20 for the heavier elements: 
$^{33}$Al \cite{Wo86} and  $^{33-35}$Si\cite{Wo85b,Fi85b,Sm86}.
On the other hand, 
the vicinity of the one-proton drip line has been explored 
by Langevin et al. \cite{La86}
by the observation $^{23}$Si, $^{27}$S, $^{31}$Ar 
and $^{35}$Ca. 
%
%
%
Recently Suzuki et al. \cite{Su96} 
have deduced matter rms radii for a series of Na and Mg isotopes 
from interaction cross sections. 

 Several shell-model calculations \cite{Wi80,Wa81,Po87,Wa90,Fu92}
have been performed for Mg isotopes.
They concluded that the inclusion of the f shell is crucial
for a correct reproduction of the properties of $^{32}$Mg. 
These studies also support the occurence of 
prolate quadrupole deformation in $^{32}$Mg. 
Until very recently\cite{Ca75},
systematic 
mean-field calculations in Na and Mg isotopes have not been numerous. 
The present work contains a drip-line-to-drip-line investigation of 
the deformation properties of the even Mg isotopes.
In addition, at the neutron 
drip line, we have also studied the neighbouring even Z elements
(Ne and Si).

\section{Strength of the pairing interaction}
 In our method, distinct parametrizations for the interaction are chosen
for the mean-field and the pairing channels.
{}For the particle-hole channel, a Skyrme force is used. 
We have studied two well established parameter sets.
The first one is SIII \cite{Be75}, which has been 
extensively and often successfully used to
study deformation properties.
The second one, Sly4 \cite{Cha96} is more recent.
It has been constructed with a special attention paid
to the properties of nuclear matter and neutron matter. 
It is therefore  expected to
provide a more realistic isospin dependence than previous forces.   
With such a choice of two parametrizations of the effective
force which are known to work equally well in the region of stable nuclei
should provide a measure of the uncertainity of microscopic 
effective theory at large $T_z$.

{}For the pairing channel, a density-dependent zero-range force is used: 
\begin{equation}
V_{\rm P}({\bf r}_1,{\bf r}_2) 
 = V_0(1-P_\sigma)(1-\frac{\rho({\bf r}_1)}{\rho_{\rm c}})
   \;\delta({\bf r}_1-{\bf r}_2) \quad,
\end{equation}
where $P_\sigma$ is the spin exchange operator and 
$\rho({\bf r})$ the total nuclear density. 
$V_0$ is the strength of the force, chosen to be the same for neutrons and
protons, and 
$\rho_{\rm c}$ is a constant which determines the density 
dependence. 
In this work we take $\rho_{\rm c}$ equal to 0.16fm$^{-3}$.
With such a value close to the nuclear saturation density,
the pairing force is strongly attractive at the nuclear surface.
This property has been discussed in \cite{Do96b} (see also ref\cite{Sta92}).
A smooth cut-off of the pairing interaction
at an energy of 5~MeV above the fermi level has been introduced following
the procedure explained in ref\cite{Bo85}.

In ref.~\cite{Te96},
we have shown that  our numerical method provides
reliable results for the properties of 
drip line nuclei when two numerial conditions are satisfied.
{}Firstly, the dimension of the box in which the HF equations are solved must
be large enough to describe correctly the tail of the wave functions. 
Secondly, since the imaginary time method used to 
solve the mean-field equations can only be applied to a limited number of
orbitals, the number of these orbitals
must be large enough to ensure a correct
description of the continuum wave functions
; especially those associated with the low-lying
resonance states.
In the present application to Mg isotopes, we have used values
for the box size (15~fm) and for the number of  single-particle wave functions
(70  for neutrons and 35 for protons) that have been shown to be
sufficient in the calculation of Ni isotopes. Since this study is devoted
to much lighter nuclei, we are confident that these 
values are adequate. 
We have also checked this point
with tests on a selection of nuclei,
which have confirmed the conclusions
of the previous study.
The Mg isotopes are not an optimal region of the nuclear chart for an
adjustment of the pairing strength $V_0$. 
Indeed, due to the weakness of pairing correlations 
in light nuclei, the determination of experimental
qp energies from binding energies is not very
accurate. 
Moreover, the variation of these binding
energies is affected by the strong dependence of the
shape  of  Mg isotopes  on the neutron number, 
which must be taken into account in the process determining
the qp energies.
Previous studies 
of superdeformation in both the A=150 and 190
mass regions have shown that 
the value $V_0$ = 1000MeVfm$^3$ for 
SIII and SLy4 gives results of good quality.
We have therefore investigated whether one could use these
values for the Mg isotopes. 
To this end, we 
have compared 
the experimental and theoretical values obtained for the
two-neutron separation energies  $S_{\rm 2n}(N,Z)$, 
calculated from the difference between
the HFB ground state energies of neighbouring even nuclei.
The quantity $-2\lambda_n$, 
where $\lambda_n$ is
the HFB chemical potential,
provides another approximation 
of $S_{\rm 2n}(N,Z)$. In order to investigate the
sensitivity of  results to $V_0$,
we have repeated some calculations
with the value $V_0$=700MeVfm$^3$.
Results are plotted on Fig.~1 together with
the experimental data. 
The $S_{\rm 2n}$ values are almost insensitive to $V_0$. 
They are also rather similar for both Skyrme parametrizations.
The general features of the experimental data are 
satisfactorily reproduced. 
Experimentally, the shell effect at N=20  disappears in $^{32}$Mg
in agreement  with the large scale shell model
calculations of ref~\cite{Fu92}. 
In this respect, the results obtained with SIII are more consistent
with the experimental data than
those obtained with SLy4, although in both cases the shell
effect is weak. The problems associated with  the
description of $^{32}$Mg will be discussed in a forthcoming publication. 

Since the strength $V_0$ = 1000MeVfm$^3$
gives a reasonnable agreement
with the experimental data for the $S_{\rm 2n}$ values,
we have decided to use it for both SIII and SLy4.
We have also checked that it leads to qp energies 
in $^{24}$Mg close to the experimental ones.
More importantly, the fact that the same value can be used
for light nuclei and for heavy superdeformed ones
is a first indication
that a unique parametrization of a 
density-dependent zero-range pairing interaction may be valid in
the whole nuclear chart.
With the adopted strength, pairing correlations
vanish for protons in $^{36-40}$Mg and for
neutrons in  $^{20}$Mg and $^{26}$Mg.

On Fig.~2, we compare the experimental data for $S_{\rm 2n}$ 
as a function of $A$ \cite{Wa93} to the prediction of different models. 
In our calculations the two-neutron drip line is located between 
$^{40}$Mg and $^{42}$Mg for both SIII and SLy4. 
The lightest Mg
isotope predicted to be bound against two proton decay is $^{20}$Mg.
Its two-proton separation energy  calculated with SIII
is equal to 3.31~MeV;
a number that is to be compared with the experimental
value 2.33~MeV. 
Other theoretical approaches designed to
reproduce nuclear masses  lead to a better agreement with 
data than our calculation. 
However, outside
the experimentally known region,
their predictions for $S_{\rm 2n}$ show irregularities the
interpretation of which is difficult. 
Nevertheless,  the trends obtained with
the relativistic mean field (RMF) theory \cite{Re96} 
and the finite-range droplet model (FRDM) \cite{M"o95}, 
are similar except for 
$^{20,22}$Mg and for a slight increase of $S_{\rm 2n}$ from 
$^{38}$Mg to $^{40}$Mg in the FRDM. 
Thus it seems that the location of the two-neutron drip line 
does not  depend very much on the theoretical approach, 
at least  for the three types of methods we have just quoted. 
Hereafter we shall discuss the results obtained with SIII unless otherwise
specified. 

\section{Deformation properties of the Mg isotopes}

{}Fig.~3 shows the variation of the energy 
of the stable Mg isotopes 
as a function of their axial quadrupole moment.
As expected, the  energy curve of $^{20}$Mg
displays the well defined spherical minimum
that one expects for the magic neutron number N=8.
In  light isotopes, there is a competition between
oblate and prolate deformations.
All deformed isotopes turn out prolate
with the exception of $^{26}$Mg.
In the heavier isotopes, a deformed shell effect appears.
It is already visible on the deformation energy curve
of $^{32}$Mg which displays an inflection point at
$Q$=1.5b.
We find well defined  prolate minima
from $^{36}$Mg to the drip line nucleus $^{40}$Mg.
In these three cases, a very shallow oblate
minimum appears at an excitation energy of 1.0~MeV. 
We therefore find that
the neutron  shell effect at $N$ = 28  is  suppressed
at the drip line. 
>From our calculations, it turns out that the onset 
of deformations at $N$=24 only generates
a drop of about 1MeV in the curve of
the evolution of $S_{\rm 2n}$ versus the neutron number.

Since $^{40}$Mg is bound by less than 2.0~MeV, 
we have checked if 
the large quadrupole moment of the ground state
was not caused by a neutron halo 
which our calculation may
not describe properly.
To test the numerical quality 
of our results concerning the space
distribution of outer neutrons, we have verified 
whether the number of neutrons
outside a sphere of large radius ($r>$15~fm)
\begin{equation}
N_{\rm out} = \int_{r\geq 15{\rm fm}} \rho_{\rm n}({\bf r}) \; d^3{\bf r}\ ,
\end{equation}
was stable against the box size $R$ 
($\rho_{\rm n}({\bf r})$ is the neutron density). 
We find $N_{\rm out}$ to be 0.430$\times 10^{-2}$ 
and 0.683$\times 10^{-2}$ 
for box sizes $R$ = 16 and 18 fm, respectively. 
These  small and almost constant values
indicate the absence of a halo in $^{40}$Mg.
In order to determine which part of the nuclear density
is responsible for the calculated  deformation,
we have introduced quantities $\beta_\tau(r_{\rm in})$ ($\tau=n,p$)
depending on 
a cutoff variable $r_{\rm c}$ :
\begin{equation}\label{e010}
 \beta_\tau(r_{\rm c}) = 
(\pi / 5)^{1/2}  \frac{
\overline{Q_\tau}(r_{\rm c})}{ \overline{r_\tau^2}(r_{\rm c})}\quad.
\end{equation}
The quantities $\overline{r_\tau^2}(r_{\rm c})$ 
and $ \overline{Q_\tau}(r_{\rm c})$ defined by
\begin{equation}
 \overline{r_\tau^2}(r_{\rm c}) =
 \int_{r\leq r_{\rm c}} r^2 \rho_\tau({\bf r}) \; d^3{\bf r}\ ,\quad
 \overline{Q_\tau}(r_{\rm c}) =
 \int_{r\leq r_{\rm c}} Q({\bf r}) \rho_\tau({\bf r}) \; d^3{\bf r}\ , 
\end{equation}
are respectively the  square radius
and  the quadrupole moment values of the fraction of the nuclear density
inside the sphere of
radius $r_{\rm c}$.
>From the results shown on Fig.~4 for $^{40}$Mg
one sees that the asymptotic values of the deformations $\beta_\tau$ are 
reached when $r_{\rm in}$ is equal to 6.0 fm.
This value is precisely equal to the 5/3rd of
the root-mean-square radius of the nucleus (3.6 fm)
and corresponds therefore to the sharp edge
liquid drop value of the outer radius.
This demonstrates that the neutron and proton deformation,
$\beta_n$ and $\beta_p$,
reflect a property of the distributions
of core nucleons.

The neutron and proton parameters $\beta_n$ and $\beta_p$ 
and the
total quadrupole moments are plotted in Fig.~5
for all the Mg isotopes.
Values of $\beta$ and of the quadrupole moments of secondary minima
are also indicated.
Due to the conjunction of the $N=Z=$12 deformed shell effects,
the nucleus $^{24}$Mg 
is the most deformed 
of the isotope chain.
In our calculation, the spherical shell closure at $N$=20 
is strong enough to compensate the
effect of the proton deformed shell closure.
On the other hand,  
we have mentioned above that the deformation energy curve
of $^{32}$Mg  [see Fig.~3] displays an inflexion
point at a prolate deformation.
This is the first indication
of the deformed shell effect which determines the shape
of the deformation energy curves of the heavier isotopes.
The values of $\beta_n$ and $\beta_p$ for this inflexion
point are reported on Fig.~5.
As the f$_{7/2}$ shell begins to be filled, 
the magnitude of the spherical neutron shell effect decreases and allows 
the proton deformed shell effect to take over in $^{36-40}$Mg.
In these nuclei, one recovers  a proton deformation similar
to that found for $^{24}$Mg.
The experimental evidence 
based on the existence of
a low energy $2^+$ state with a large B(E2) value \cite{Mo95} in 
$^{32}$Mg implies that our calculation overestimates
the magnitude of the spherical N=20 shell effect in this nucleus.
With two other Skyrme parametrizations, Sly4
and SkP \cite{Do84}, we also find that the deformation energy 
curve for $^{32}$Mg presents a spherical minimum.
To our knowledge, this is also the case
for all mean-field calculations whether 
they include a treatment of pairing correlations\cite{Po87} or not.
Apart from $^{32}$Mg, data on deformation
are few with generally large error bars.
Still they are consistent with values of $\beta$ 
larger than those  we have calculated.
We believe that this systematic effect reflects 
the approximate character
of the the prescription (\ref{e010}) that we
use to define $\beta_p$.
Indeed, it is implicitly based on a sharp edge liquid
drop picture which neglegts surface effects which could
be of importance in such light nuclei.
Keeping this restriction in mind,
we note that for light isotopes, the values of $\beta_n$ and $\beta_p$
are similar, while
they differ for the last three heavy isotopes.
With mean-field effective forces, 
differing deformations are more easy to obtain when there is
a large excess of neutrons over protons.
Indeed, for nuclei close
to the stability line,
the large overlap between the neutron and proton densities 
and the strongly attractive 
neutron-proton interaction have a tendancy to 
suppress the differences of deformations.
This is indeed what we find for the light magnesiums.
In view of our results, it appears worthwhile to
experimentally investigate whether heavy Mg isotopes would not be 
the lightest examples of nuclei with
differing neutron and proton deformations.

On Fig.6, one sees that the calculated root-mean-square radii  
stay within the error bars
of the measured data. 
The onset of deformation in
$^{36}$Mg causes  a slight irregularity in the calculated curves. 
An irregularity of the same magnitude
does not seem to be present in the experimental data around $^{32}$Mg.

The single particle
level energies in the canonical Hartree-Fock-Bogoliubov basis are shown
on Fig.~7 for $^{40}$Mg.
Apart from a global shift in energy, a similar diagram
is valid for all Mg isotopes.
In $^{40}$Mg, at the quadrupole moment of the deformed ground state,
the neutron fermi level is close to the continuum.
The large density of levels
at the fermi surface 
provides an explanation for the smaller
neutron deformation (compared to proton).
Indeed, because it enhances the effect of pairing correlations, 
it induces a population of orbitals with a less deformation
driving character than the $\pi$d$_{5/2}$ orbital.

Results qualitatively similar to ours have been obtained in the RMF
calculation of Ren et al.\cite{Re96}. 
Both $^{20}$Mg and
$^{32}$Mg have been found spherical by these authors and
$^{36-40}$Mg very deformed. 
However, the quadrupole
moments of deformed isotopes are significantly
larger in the RMF calculation. This may be related
to the  BCS approximation with a constant gap
used by Ren et al.

The $N$ = 28 neutron gap does show up as an accident
neither in the $S_{\rm 2n}$ curve nor in the 
evolution of the radii.
On Fig. 7, one sees that the value of the $N$=28 single particle gap is
about 3 MeV in $^{40}$Mg.
This is much smaller than the gap 
that one finds in $^{48}$Ca (5MeV)
with a calculation using the same effective force\cite{Be75}.
This quenching of the $N$ = 28 gap  
is sufficient to allow the deformed proton gap at $Z$ = 12
to win over the spherical shell effect and to 
lead to a deformed nucleus.
Similar results have been obtained in a recent study of 
Si and S isotopes \cite{We96}.

 Fig.~8  shows the potential energy curve of
the nucleus $^{42}$Si ($N$ = 28).
 The same figure
displays the curves of the two nuclei $^{34}$Ne and $^{46}$Si
which according to our method are on the drip line.  
The energy curves of these three isotopes display
a soft dependence 
on the quadrupole moment. 
This suggests that dynamical quadrupole collective
effects may play an important r\^ole.
One notes also that, as was the case already for $^{40}$Mg,
there is no indication of a spherical shell 
effect at $N$ = 28 for $^{42}$Si.
Similar results have been obtained by Werner et al\cite{We96}
with the HF+BCS approximation.

\section{Conclusion}

In this paper, we have reported the first calculation of 
the deformation properties of
drip line nuclei using a Hartree-Fock-Bogoliubov
approach in which pairing correlations in the
continuum are correctly
treated. 

Our work shows that deformation effects
cannot be neglected 
in a description of the structure 
of neutron rich Mg isotopes.
In particular, it provides a negative answer to the
question raised by Tanihata et al \cite{Ta95} as to
whether all drip line nuclei are spherical. 

Another conclusion of this work, is that the N=28 spherical shell closure 
is significantly weakened
in light nuclei close to the drip line. 
As a consequence, we find that the deformation properties
of heavy Mg isotopes are more influenced by
the proton deformed shell effect at $Z$=12 than by the
neutron spherical shell closure at $N$=28. 
Due to pairing correlations, the
population of neutron deformation
driving orbits favours 
the occurence of deformation in the last bound Mg isotopes. 

{}Finally, we find that
with the prescription that we introduce to define deformations,
the large excess of
neutrons with respect to the protons leads
non zero differences ($\beta_n-\beta_p$)
in the heavy Mg isotopes.
A confirmation of this property would be the
existence of low lying quadrupole isovector excitation
for isotopes close to the drip lines.

Therefore, the several original
properties predicted by our calculation give an additional incentive
to the pursuit of a more thorough investigation 
of this section of neutron drip line.

\section*{Acknowledgments}
This research 
 was supported in part 
 by the Belgian Office for Scientific Policy under Contract ARC~93/98-166.

\newpage
\noindent
{}Figure Captions

\begin{itemize}
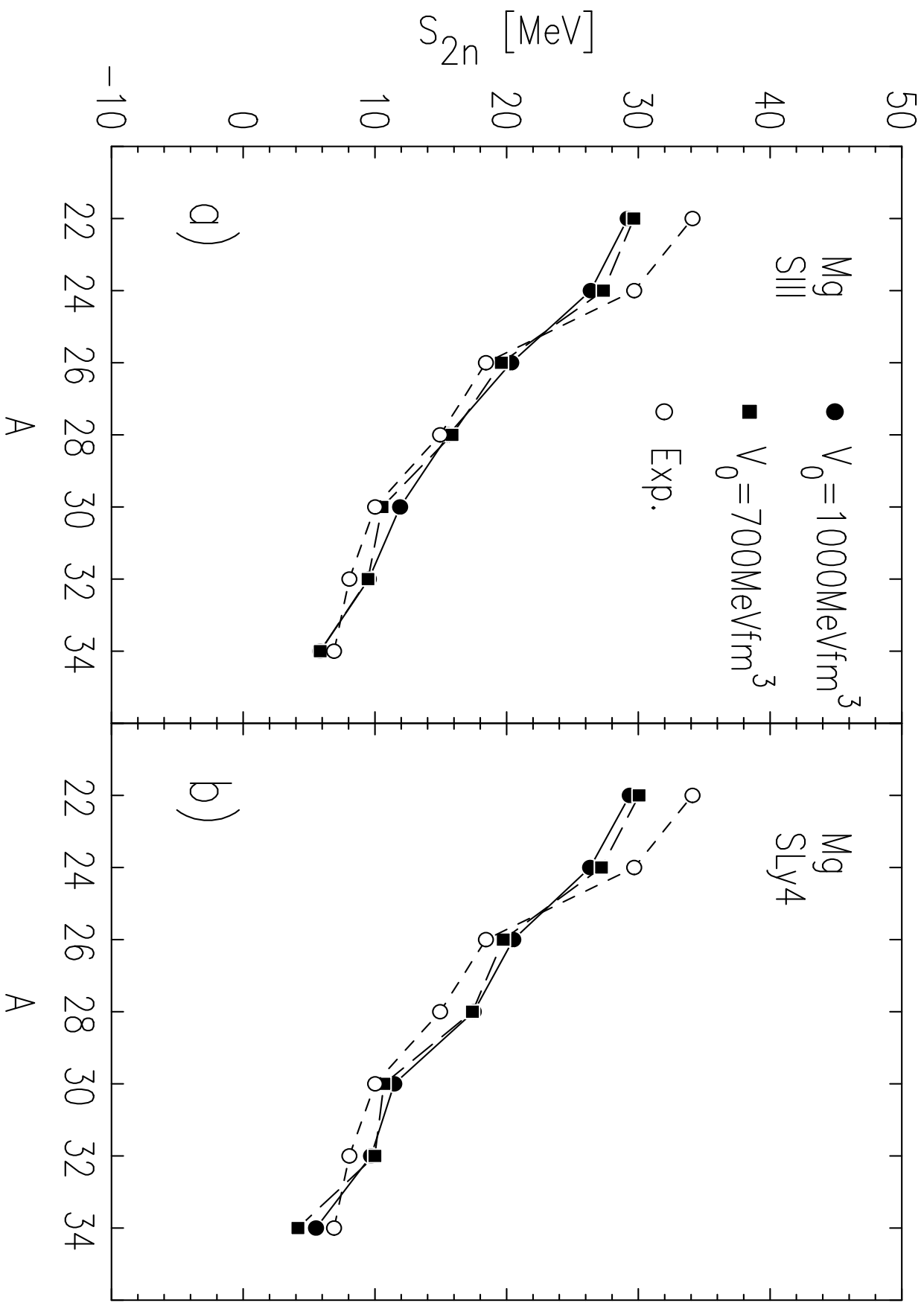
\item[Fig.~1] Calculated and experimental $S_{\rm 2n}$ of Mg
isotopes for 
$A$ = 22$-$34. 
with two Skyrme force parametrizations: (a) SIII and (b) SLy4.
In both cases, two strengths of the pairing force are
tested: $V_0$ = 1000 and 700 MeV$\,$fm$^3$. 
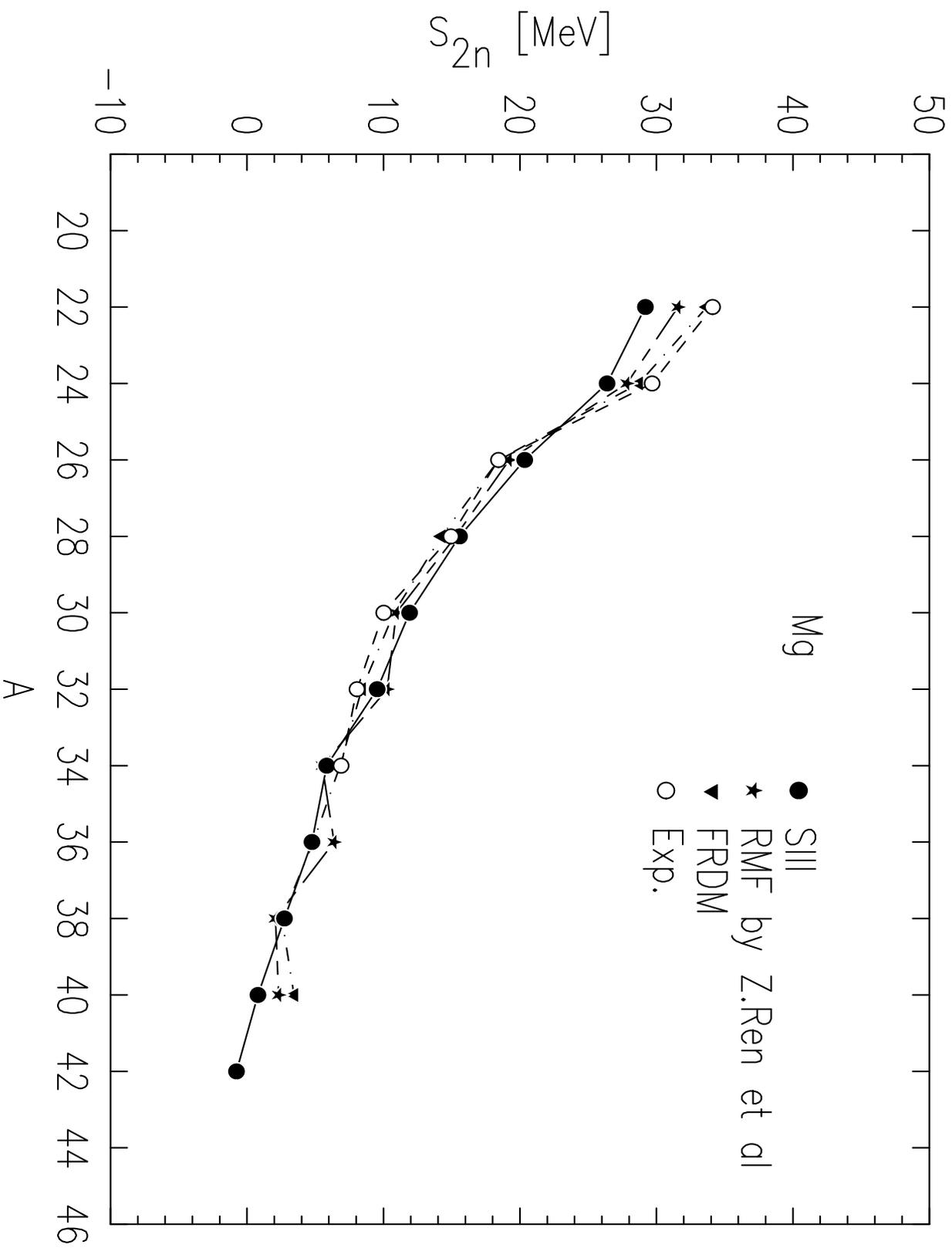
\item[Fig.~2] Calculated and experimental $S_{\rm 2n}$
from the proton to the neutron drip lines. 
Our results are obtained with the SIII parametrization.
Results of RMF and FRDM calculations are taken from 
refs.~\cite{Re96} and \cite{M"o95}, respectively. 
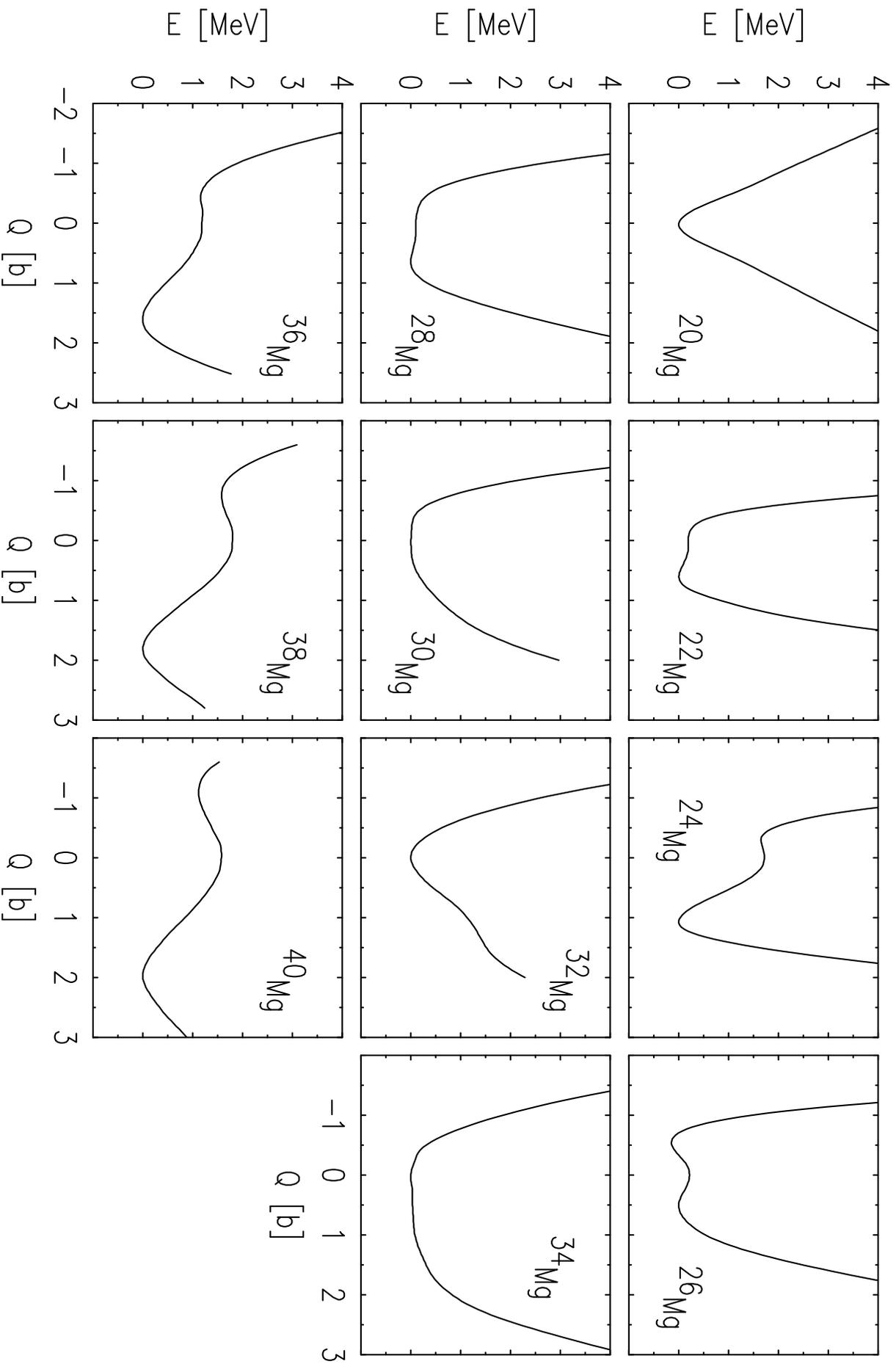
\item[Fig.~3] Deformation energy curves calculated for 
the even $^{20-40}$Mg
isotopes with SIII as a function of the axial  quadrupole moment. 
The origin of the energy is taken at the minimum of each curve. 
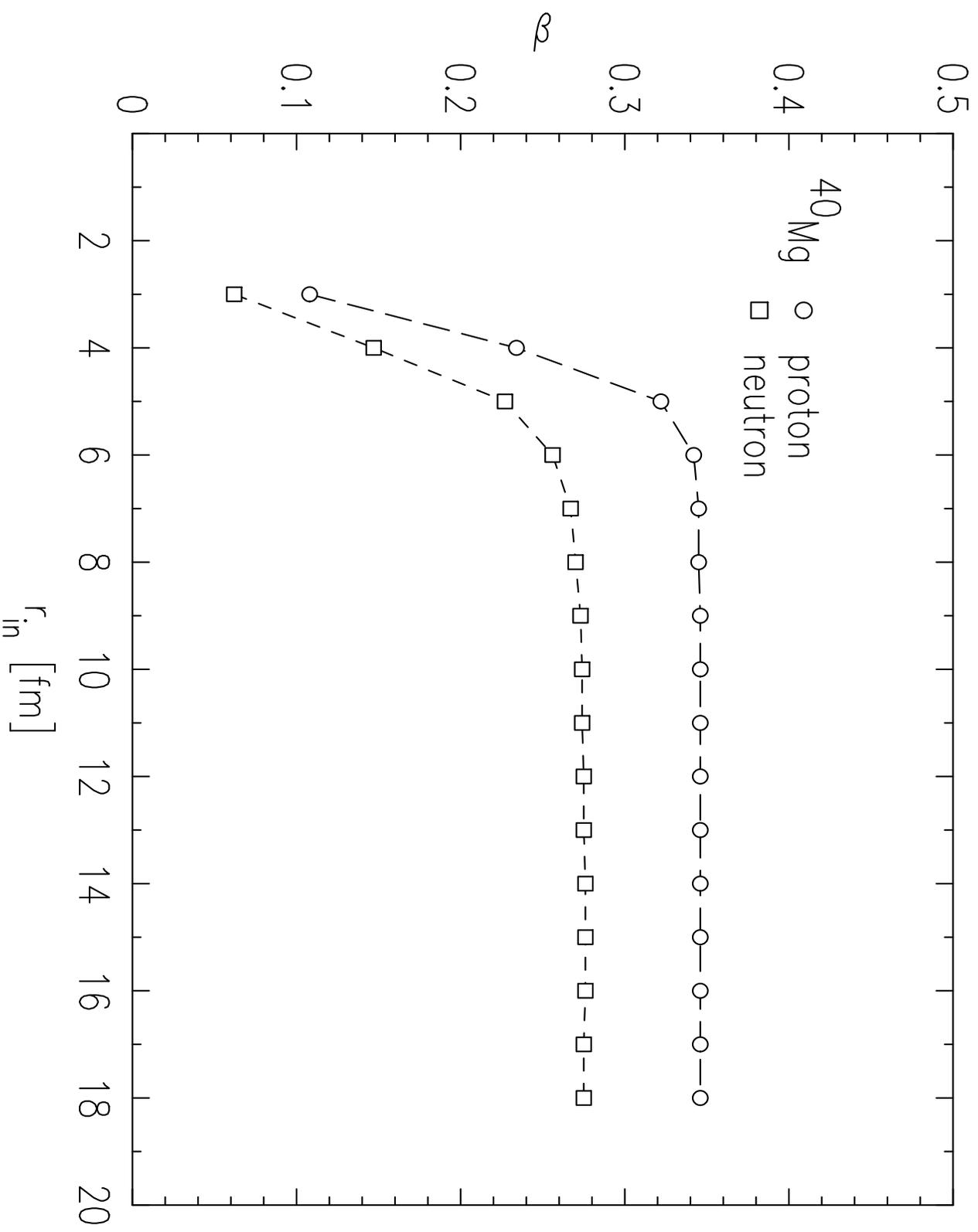
\item[Fig.~4] Deformation parameter $\beta$ 
for protons and neutrons in $^{40}$Mg 
calculated as a function of the cut-off radius $r_{\rm in}$.
See eqs.~(3) and (4). 
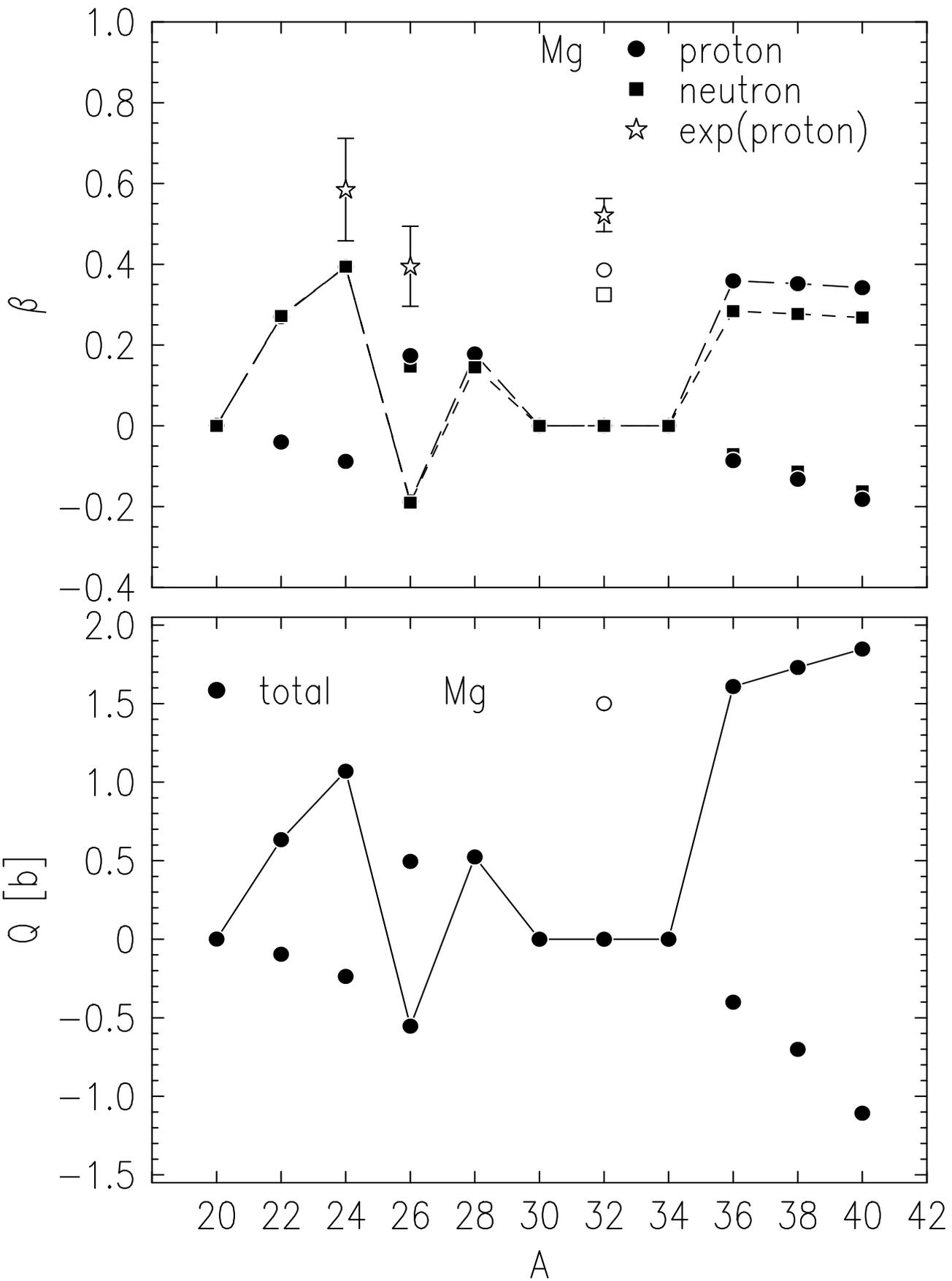
\item[Fig.~5] Deformation parameter $\beta$ (upper panel) 
and total quadrupole moments (lower panel) of the Mg isotopes
versus $A$. 
Ground state values are connected by lines.
Isolated symbols correspond 
to the secondary minima found for some isotopes. 
Black symbols correspond to absolute or secondary
minima, open symbols to shoulder in the deformation
energy curves of Fig. 3.
The $\beta$ deduced from the experimental data have
been taken from refs.~\cite{Sc72} 
and \cite{Le76} for $^{24,26}$Mg  and from \cite{Mo95}
for $^{32}$Mg. 
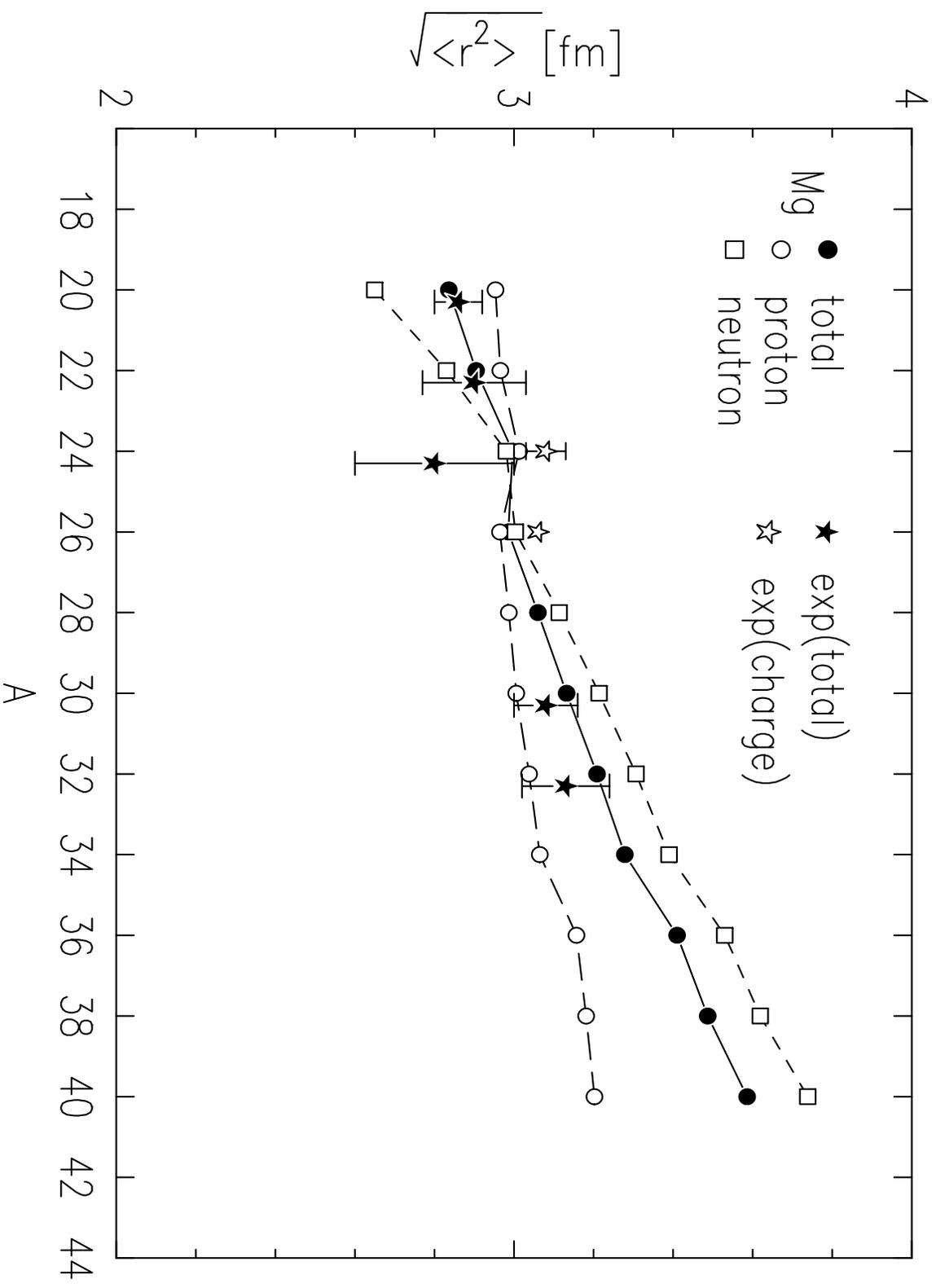
\item[Fig.~6] Calculated and experimental rms radii versus $A$. 
The theoretical values are joined by lines.
The black stars are the matter rms radii of ref.\cite{Su96}, and
the white stars 
are the charge rms radii of ref.\cite{Le76}. 
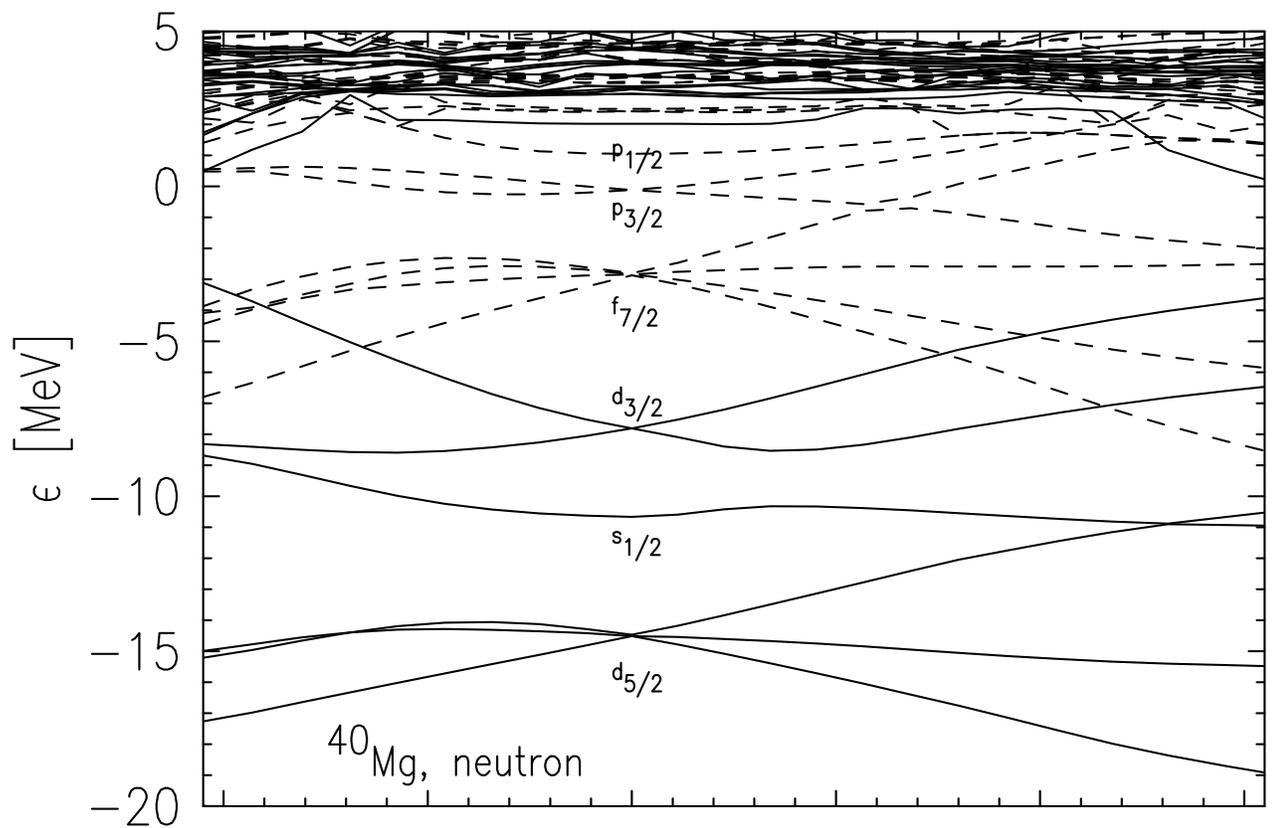
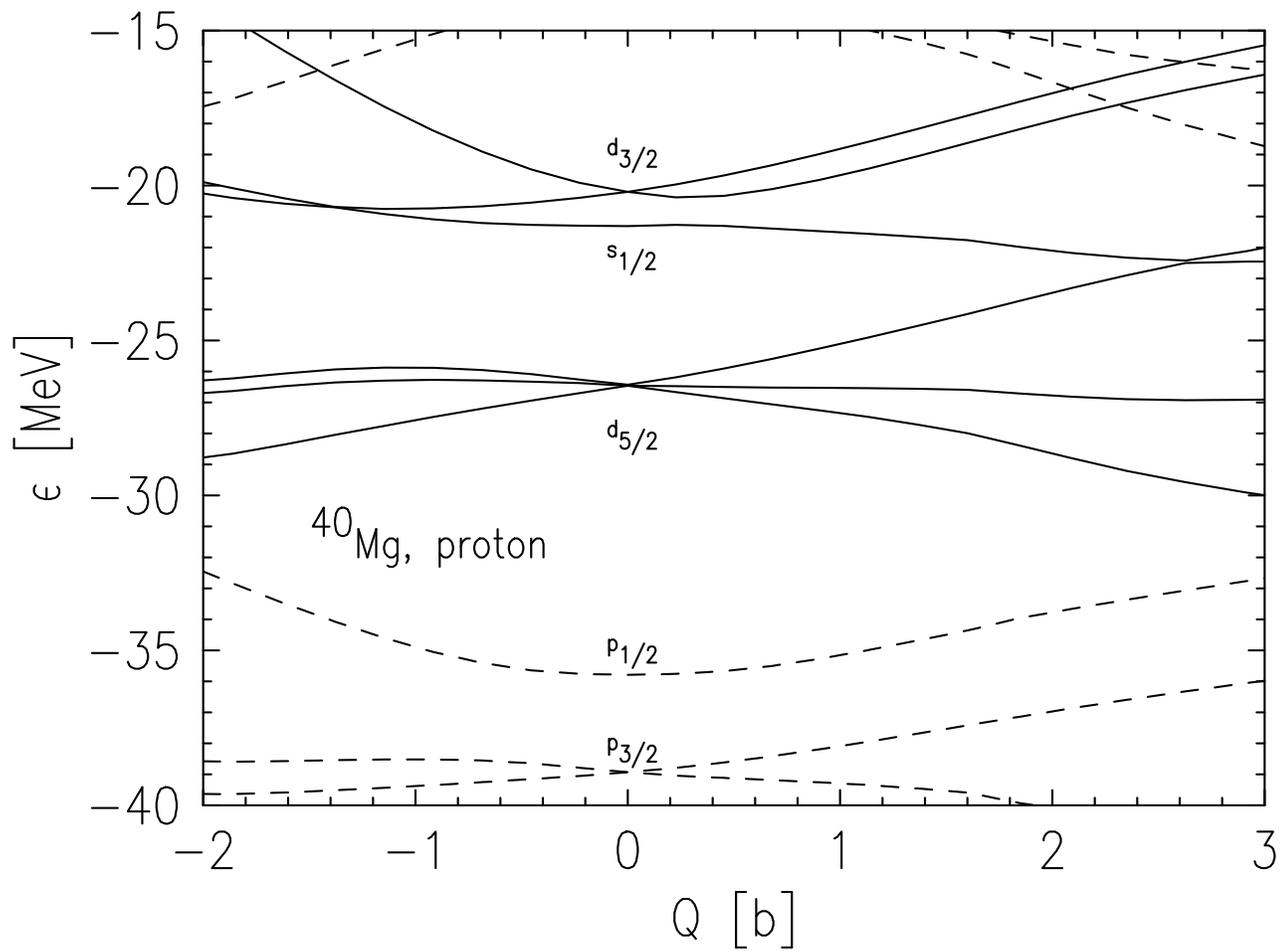
\item[Fig.~7] Single-particle energy diagram of $^{40}$Mg as a function 
of the total quadrupole moment. 
These energies are 
the diagonal matrix elements of the HF Hamiltonian in the canonical basis. 
The upper and lower parts are for neutrons and protons respectively. 
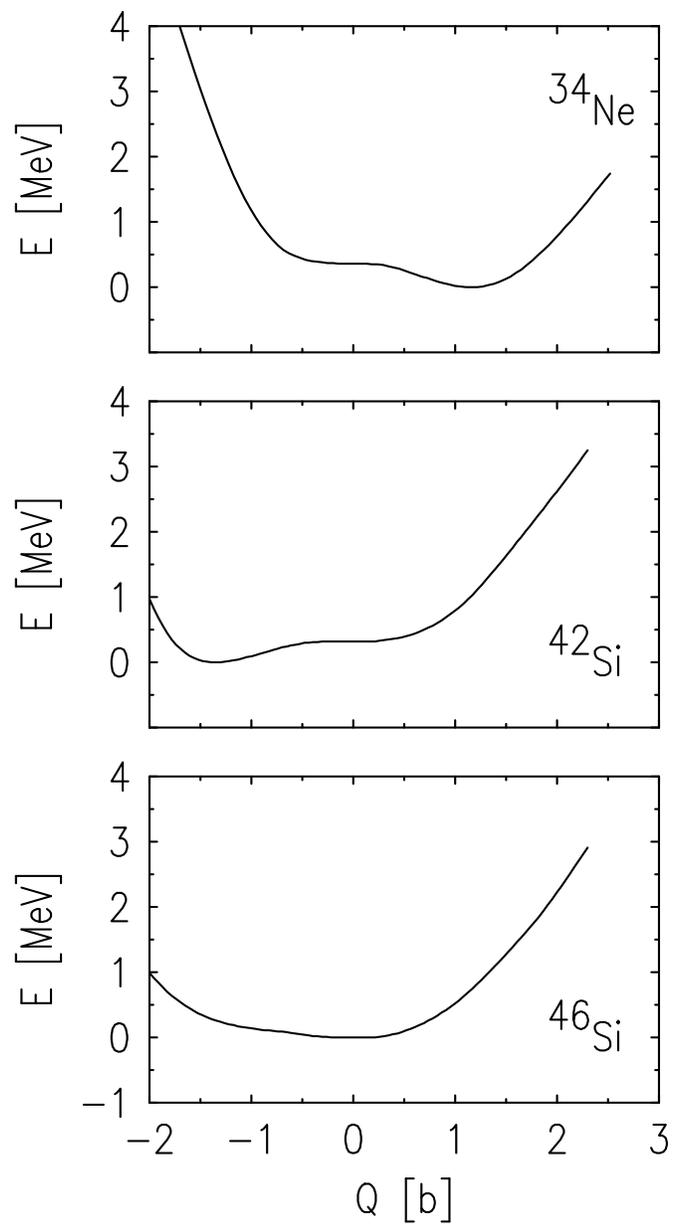
\item[Fig.~8] Deformation energy curves of $^{34}$Ne and $^{42,46}$Si
as a function of the axial quadrupole moment. 
\end{itemize}
\end{document}